\documentclass[letterpaper,12pt]{article} \usepackage{amsmath,amsfonts}             \normalsize \def\TC#1{} \def\OC#1{#1} \def\BL{}
\newcommand{\OCTC}[2]{\OC{#1}\TC{#2}}  
\newcommand{\LBS}{\OCTC{\left[}{\bigg[}}
\newcommand{\RBS}{\OCTC{\right]}{\bigg]}}
\newcommand{\LBR}{\OCTC{\left(}{\bigg(}}
\newcommand{\RBR}{\OCTC{\right)}{\bigg)}}
\newcommand{\LBB}{\OCTC{\left\{}{\bigg\{}}
\newcommand{\RBB}{\OCTC{\right\}}{\bigg\}}}
\newcommand{\Ds}{D\!\!\!\!/}
\newcommand{\dels}{\partial\!\!\!/}
\newcommand{\dotp}{\cdot}
\newcommand{\porm}{\pm} 

\newcommand{\boldg}{\boldsymbol{\gamma}}
\newcommand{\kiok}{\frac{k_i}{k}}
\newcommand{\kjok}{\frac{k_j}{k}}
\newcommand{\tilpsii}{{\tilde{\psi_i}}}
\newcommand{\kvec}{{\bf k}}
\newcommand{\xvec}{{\bf x}}
\newcommand{\yvec}{{\bf y}}
\newcommand{\kdg}{\kvec \dotp \boldg}
\newcommand{\half}{\frac{1}{2}}
\newcommand{\kdgok}{\frac{\kdg}{k}}
\newcommand{\mpo}{\frac{m_+}{d-1}}
\newcommand{\gdb}{\boldg \cdot {\bf b}^+}
\newcommand{\kdbok}{\frac{{\bf k} \cdot {\bf b}^+}{k}}
\def\Pcurlymath#1{{\mathcal{#1}}}

\title{\OC{The} \BL{\LARGE} AdS/CFT Correspondence for the Massive Rarita-Schwinger Field}
\author{\BL
	P.~Matlock\thanks{e-mail: \texttt{pwm@sfu.ca}}\ ~and
	K.~S.~Viswanathan\thanks{e-mail: \texttt{kviswana@sfu.ca}}\\ \small Department
	of Physics, Simon Fraser University, Burnaby, B.C., Canada
	}

\begin{document} 
\maketitle 
\begin{abstract} 
\BL
The complete solution to the massive Rarita-Schwinger field equation in anti-de~Sitter space
is constructed, and used in the AdS/CFT correspondence to calculate the correlators for the boundary
conformal field theory. It is found that when no condition is imposed on the field 
solution, there appear two different boundary conformal field operators, one coupling to a 
Rarita-Schwinger field and the other to a Dirac field. These two operators are seen to
have different scaling dimensions, with that of the spinor-coupled operator exhibiting 
non-analytic mass dependence.
\end{abstract} 

\newcommand{\alphasign}{+} 

\section{Introduction} 
\label{INTROsection}
\BL
The Maldacena conjecture \cite{Maldacena} asserts that there exists 
a holographic correspondence \cite{Susskind} between
field theories on $d+1$-dimensional AdS space and conformal field theories
on the $d$-dimensional boundary of this space. This correspondence has been made more precise in
\cite{Maldacena,GubsKlebPolya,ArefevaVolovich,Witten,KlebWitt} and investigated for specific cases in
\TC{\cite{MuckVish0,HenningsonSfestos,MuckVish1,DHokerFreeSki,Tseytlin,Corley,Volovich,FreedmanMathur,MuckVish2,KoshelevRitchkov,Rashkov}.} 
\OC{\cite{MuckVish0}-\cite{Rashkov}.}
We do not attempt to give a comprehensive 
list of references here, but refer the reader to the literature. A recent 
review of the Maldacena conjecture can be found in \cite{Petersen}.
According to this correspondence principle, the action for the field theory in the bulk AdS space
written in terms of the boundary values of the fields serves as a generating functional for a field
theory which lives on the flat boundary space. This can be written
	\begin{equation}
        \label{Malda}
          Z_{AdS}[\psi_{(0)}]=\int_{\psi_{(0)}} \Pcurlymath{D} \psi e^{-I[\psi]} \TC{\cr}
		= Z_{CFT}[\psi_{(0)}] = \left\langle e^{\int_{\partial\textup{AdS}}
			\textup{d}^dx \Pcurlymath{O} \psi_{(0)}} \right\rangle
	\end{equation}
where $\psi_{(0)}$ is the boundary field, and acts as a source for the operator $\Pcurlymath{O}$.
Since we deal in the present case with a classical field, we obtain an approximation to the path 
integral.\footnote{An investigation of the AdS/CFT correspondence which deals with quantum corrections in given in \cite{DHokerFreeSki}} 

We will choose the AdS metric to 
be $g_{\mu\nu}=\frac{1}{{x^0}^2}\delta_{\mu\nu}$ so that the boundary is at $x^0=0$. Since the metric 
diverges on this boundary, we must regularise by multiplying by a function with a suitable zero on 
the boundary. \cite{Witten} The fact that this function is otherwise unspecified is the origin of the 
conformal invariance in the boundary field theory.

Of particular interest here are \cite{Corley,Volovich,KoshelevRitchkov,Rashkov} which also
deal with the Rarita-Schwinger field, but impose restrictions on the solution of the field equation.
We construct the general solution and find that when no such restrictions are imposed, there 
appear two fields on the $d$-dimensional 
boundary; both a Dirac spinor and a spin $3/2$ Rarita-Schwinger field couple to boundary
conformal field operators, which as a result have different conformal scaling dimensions.
To find these conformal field correlators, we use the Dirichlet boundary value problem method 
exhibited in, for example, \cite{MuckVish1} for the case of a Dirac spinor field. Since the action
vanishes on-shell, a surface term must be added.\footnote{The equations of motion, of course, do not change.}
Two equivalent methods \cite{ArutyunovFrolov,Henneaux} of determining this term have been investigated,
and the method of \cite{Henneaux} has recently been used in \cite{Rashkov}. 

In following this prescription, we solve the equations of motion in section \ref{EOMsection}. The surface term
to add to the action is found using the method of \cite{Henneaux} in section \ref{SURFsection}, and finally 
in section \ref{CORRsection} the CFT correlators are calculated. These correlators are fixed, up to a 
multiplicative factor, by conformal invariance \cite{FerGrilPari,Zaikov}. The results obtained are 
consistent with these considerations.

\section{Solving the Classical field Equations} 
\label{EOMsection}
Although the equations of motion have been solved in \cite{Corley} for the 
massless case, in \cite{Volovich} for the case of $\gamma^\mu\psi_\mu=0$, and 
somewhat more generally in \cite{KoshelevRitchkov}, we find it necessary
to construct explicitly the complete solution to the massive\footnote{As pointed out 
in \cite{KoshelevRitchkov}, one
must consider $m_1 \neq 0$ in the case of supergravity on $AdS_5\times S^5$ \cite{vanN}.}
case while imposing no restrictions. 

Our index conventions are  $\mu,\nu,... = 0 ... d$ and $i,j,... = 1 ... d$.
We choose the metric of $AdS_{d+1}$ to be $g_{\mu\nu}=\frac{1}{{x^0}^2}\delta_{\mu\nu}$ so 
that $AdS$ space is given by $x^0>0$. 
The boundary with which we shall be concerned is at $x^0=0$, where the metric is singular.
The Rarita-Schwinger action is given by
	\begin{equation}
	\label{RSaction}
	I=\int d^{d+1}x \sqrt{g}\overline{\psi}_\mu[\Gamma^{\mu\nu\sigma}D_\nu -
		 m_1 g^{\mu\sigma} - m_2 \Gamma^{\mu\sigma}] \psi_\sigma
	\end{equation}
$D_\nu$ denotes the covariant derivative, $\Gamma_\mu$ are curved space Dirac matrices so that
$\Gamma_\mu=e_\mu^a\gamma_a$ where $\{\gamma_a,\gamma_b\}=2\delta_{ab}$ are Euclidean Dirac 
matrices, which are taken to be Hermitian, and the vielbein is given by $e^a_\mu=\frac{1}{x^0}\delta_\mu^a$. More than one index on 
these matrices indicates antisymmetrisation (including $1/n!$). 
Varying the above action gives the Rarita-Schwinger equation
	\begin{equation}
	\label{1.1}
	[\Gamma^{\mu\nu\sigma}D_\nu - m_1 g^{\mu\sigma} - m_2 \Gamma^{\mu\sigma}] \psi_\sigma=0
	\end{equation}
and its conjugate
	\begin{equation}
	\label{1.1conj}
	\overline{\psi}_{\mu}[\Gamma^{\mu\nu\sigma}\overleftarrow{D}_\nu + m_1 g^{\mu\sigma} 
		+ m_2 \Gamma^{\mu\sigma}] =0
	\end{equation}
We will find it convenient to write the former in the equivalent form
	\begin{equation}
	\Gamma^\nu[D_\nu\psi_\mu-D_\mu\psi_\nu]+\mpo\Gamma_\mu\Gamma^\nu\psi_\nu-m_-\psi_\mu=0
	\end{equation}
which can be seen by using
$\Gamma^{\mu\nu\sigma}=\half(\Gamma^\nu\Gamma^\sigma\Gamma^\mu - \Gamma^\mu\Gamma^\sigma\Gamma^\nu$).
To solve this equation generally, we first contract with $D_\mu$ to obtain
	\begin{equation}
	\label{1.2}
	\Gamma^{\mu\nu\sigma}[D_\mu,D_\nu]\psi_\sigma - 2m_1 D^\mu \psi_\mu -
	 m_2 \Gamma^{\mu\nu}D_{[\mu}\psi_{\nu]}=0
	\end{equation}
The commutator $[D_\mu,D_\nu]\psi_\sigma$ can be expressed as
	\begin{equation}
	\label{3.1}
	[D_\mu,D_\nu]\psi_\sigma =\left( \frac{1}{2} \partial_{[\mu}\omega_{\nu]}
	+ \frac{1}{4} [\omega_\mu,\omega_\nu] \right)\psi_\sigma = \half R_{\mu\nu}\psi_\sigma
	\end{equation}
where the spin connection is given by
	\begin{equation}
 	\omega_\mu^{AB}=\frac{1}{x^0}(\delta^A_0\delta^B_\mu-\delta^B_0\delta^A_\mu)
	\end{equation} 
and $\omega_\mu=\omega_\mu^{AB}\Sigma_{AB}$. The computation of $R_{\mu\nu}$ is simplified if we make 
use of the fact that the space is maximally symmetric \cite{Weinberg}. We find
	\begin{equation}
	R_{\mu\nu}=\frac{R}{2d(d+1)}[\Gamma_\mu,\Gamma_\nu] 
	\end{equation}
and in our metric $R=-d(d+1)$ so that
	\begin{equation}
	\label{4.3}
	\Gamma^{\mu\nu\sigma}[D_\mu,D_\nu] \psi_\sigma = \frac{d(d-1)}{2} \Gamma^\sigma \psi_\sigma
	\end{equation}
and \eqref{1.2} becomes
	\begin{equation}
	\label{1.2becomes}
	m_2 \Ds (\Gamma^\mu\psi_\mu) +  m_- D^\mu \psi_\mu  + \frac{d(d-1)}{4}\Gamma^\mu\psi_\mu  =0
	\end{equation}
Now we contract \eqref{1.1} with $\Gamma_\mu$. 
Using $\Gamma_\mu\Gamma^{\mu\nu\sigma}=(d-1)\Gamma^{\mu\nu}$ we find
	\begin{equation}
	\label{2.5}
	\Ds (\Gamma^\mu\psi_\mu) - D^\mu\psi_\mu + \frac{m_1+dm_2}{1-d}\Gamma^\mu\psi_\mu = 0
	\end{equation}
Combining \eqref{1.2becomes} and \eqref{2.5} to eliminate $D^\mu\psi_\mu$, we obtain a Dirac equation 
	\begin{equation}
	\label{4.4}
	[\Ds-C](\Gamma^\nu\psi_\nu)=0
	\end{equation}
where
	\begin{equation}
	\label{10.4}
	C=\frac{d(d-1)}{4m_1}+\frac{(m_1+dm_2)m_-}{m_1(d-1)}
	\end{equation}
and for convenience we have defined $m_{\porm} = m_1 \porm m_2$.
It can also be shown from \eqref{1.2becomes} and \eqref{2.5} that $m_1=0$ implies
$\gamma^\mu\psi_\mu=0$. Since this case has been considered in \cite{Corley} and \cite{Volovich},
we assume $m_1 \neq 0$.   

Now we specialise to our coordinate system and write \eqref{4.4} as
	\begin{equation}
	\label{5.2}
	\left(x^0\dels-\frac{d}{2}\gamma_0-C\right)\gamma \dotp \psi=0
	\end{equation}
where we will now work only with the components $\psi_a \equiv e_a^\mu\psi_\mu$.

This equation has been solved in \cite{MuckVish1} by differentiating to obtain a second-order equation, and 
has the solution which does not diverge as $x^0 \rightarrow \infty $  	
	\begin{equation}
	\label{8.1psi}
	\gamma \cdot \psi=(kx^0)^\frac{d+1}{2}
			\left[
		 	A^{(1)} K_{C + \frac{1}{2}}(kx^0) +  A^{(3)} K_{C - \frac{1}{2}}(kx^0)
			\right]
	\end{equation}
where $A^{(1)}$ and $A^{(3)}$ are spinors which do not depend on $x^0$.
Since this form of the solution was found via a second order equation, 
equation \eqref{8.1psi} needs to be substituted back into the first-order equation \eqref{5.2}
in order to find these spinors.
We write $x=(x^0,{\bf x})$, and we will work in Fourier space with respect to the non-zero-index
components of the field,
	\begin{equation}
	\label{fourier}
	\tilde{\psi}_\mu(x^0,\kvec)=\int d^dx e^{i \kvec \cdot \xvec }\psi_\mu(x)
	\end{equation}
Since we will soon need to work with several other first-order equations and doing the full calculation
every time would be tedious, we calculate the following formula
 \begin{eqnarray}
 \label{9.1}
\OCTC{
	 \lefteqn{\left[x^0\gamma_0{\partial}_0-ix^0{\kdg}-n\gamma_0-P\right](kx^0)^l 
		\left[\big(A^{(1)}
		+(kx^0)A^{(2)}\big)K_{q+\frac{1}{2}}
		+\big(A^{(3)}+(kx^0)A^{(4)}\big)K_{q-\frac{1}{2}}\right] } \nonumber\\
 	 &=& (kx^0)^l \Bigg\{ K_{q+\half}\left[(l-n-q-\half)\gamma_0-P\right]A^{(1)}
	         +K_{q-\half}\left[(l-n+q-\half)\gamma_0-P\right]A^{(3)} \nonumber\\
         &+& (kx^0)K_{q+\half}\left[-i\kdgok A^{(1)} +\big((l-n-q+\half)\gamma_0-P\big)A^{(2)}
		-\gamma_0A^{(3)}\right] \nonumber\\
	 &+& (kx^0)K_{q-\half}\left[-i\kdgok A^{(3)} +\big((l-n+q+\half)\gamma_0-P\big)A^{(4)}
	 	-\gamma_0A^{(1)}\right] \nonumber\\
         &+& (kx^0)^2K_{q+\half}\left[-i\kdgok A^{(2)} -\gamma_0A^{(4)}\right] 
	     +(kx^0)^2K_{q-\half}\left[-i\kdgok A^{(4)} -\gamma_0A^{(2)}\right] \Bigg\}
}{   
	 \lefteqn{\left[x^0\gamma_0{\partial}_0-ix^0{\kdg}-n\gamma_0-P\right](kx^0)^l} \nonumber\\
		&\times& \left[\big(A^{(1)}
		    +kx^0A^{(2)}\big)K_{q+\frac{1}{2}}
		    +\big(A^{(3)}+kx^0A^{(4)}\big)K_{q-\frac{1}{2}}\right]  \nonumber\\
 	 &=& (kx^0)^l \Bigg\{ K_{q+\half}\left[(l-n-q-\half)\gamma_0-P\right]A^{(1)} \nonumber\\
	 &+& K_{q-\half}\left[(l-n+q-\half)\gamma_0-P\right]A^{(3)} \nonumber\\
         &+& (kx^0)K_{q+\half}\bigg[-i\kdgok A^{(1)} +\big((l-n-q+\half)\gamma_0 \nonumber\\
	 &-& P\big)A^{(2)}-\gamma_0A^{(3)}\bigg] \nonumber\\
	 &+& (kx^0)K_{q-\half}\bigg[-i\kdgok A^{(3)} +\big((l-n+q+\half)\gamma_0 \nonumber\\
	 &-& P\big)A^{(4)}-\gamma_0A^{(1)}\bigg] \nonumber\\
         &+& (kx^0)^2K_{q+\half}\left[-i\kdgok A^{(2)} -\gamma_0A^{(4)}\right] \nonumber\\
	 &+& (kx^0)^2K_{q-\half}\left[-i\kdgok A^{(4)} -\gamma_0A^{(2)}\right] \Bigg\}
	}
\end{eqnarray}
where $P,l,n,q$ are arbitrary constants, the $A$ are spinors which may depend on $\kvec$, and 
from now on we omit the arguments $(kx^0)$ of the Bessel functions. 
Our present case of $\gamma \cdot \tilde{\psi}$ corresponds to \eqref{9.1} with 
$A^{(2)}=A^{(4)}=0$ and $q=P=C, n=\frac{d}{2}, l=\frac{d+1}{2}$. 
Requiring the resulting RHS of \eqref{9.1} to vanish gives $A^{(1)}$ and
 $A^{(3)}$, so that, writing $k=|\kvec|$,
	\begin{equation}
	\label{10.3}
	\gamma \cdot \tilde{\psi}=(kx^0)^{\frac{d+1}{2}}
	\left[
	i\kdgok K_{C + \frac{1}{2}} 
	 +  K_{C - \frac{1}{2}}
	\right]b_0^+(\kvec)
	\end{equation}
where $b_0^+$ is a free spinor function of $\kvec$.
(We will consistently use $+$ and $-$ superscripts to denote eigenspinors of $\gamma_0$ with eigenvalues $+1$ and $-1$.)

The equation of motion \eqref{1.1} may be written
	\begin{equation}
	\label{12.1}
	\left[x^0\gamma \cdot {\partial} - \frac{d}{2}\gamma_0-m_-\right]\psi_a \TC{\cr}
		+\left[\frac{3}{2}\delta_{a0}-
		x^0{\partial}_a-\frac{1}{2}\gamma_0\gamma_a+\mpo\gamma_a\right]\gamma \cdot \psi 
	= \gamma_a\psi_0
	\end{equation}
The $a=0$ component of \eqref{12.1} is
	\begin{equation}
	\label{12.2}
	\left[x^0\gamma_0{\partial}_0-ix^0{\kdg}-(\frac{d}{2}+1)\gamma_0-m_-\right]\tilde{\psi_0} \TC{\cr}
	=\left(x^0{\partial}_0-1-\mpo\gamma_0\right)\gamma\cdot\tilde{\psi}
	\end{equation}

We will find both a particular and homogeneous solution for \eqref{12.2}.
Since we already know the RHS, we make the ansatz
	\begin{equation}
	\label{13.1}
	\tilde{\psi_0}^P=(kx^0)^\frac{d+1}{2}
		\LBS
		\big(A^{(1)}+(kx^0)A^{(2)}\big)K_{C+\frac{1}{2}} \TC{\cr}
		+\big(A^{(3)}+(kx^0)A^{(4)}\big)K_{C-\frac{1}{2}}
		\RBS
	\end{equation}
(Note that we re-use the parameters $A^{(1)},A^{(2)},A^{(3)},A^{(4)}$ each time
we do a calculation with \eqref{9.1}.)
Reading off the LHS of \eqref{12.2} as the RHS from \eqref{9.1} with
$P=m_-,q=C,n=\frac{d}{2}+1,l=\frac{d+1}{2}$ and matching the coefficients of the linearly 
independent functions of $kx^0$ with those on the RHS of \eqref{12.2}, we obtain equations 
which can be solved for the $A$ parameters. The result is
	\begin{equation}
	\label{14.1}
	\tilde{\psi_0}^P=(kx^0)^\frac{d+1}{2}
	\LBB
	K_{C+\half}\left[i\mu_2\kdgok + \mu_3(kx^0)\right]b_0^+ \TC{\cr}
	+K_{C-\half}\left[-\mu_1+i\mu_3(kx^0)\kdgok\right]b_0^+
	\RBB
	\end{equation}
where
	\begin{equation}
	\label{14.1mu}
	\mu_1=\frac{\mpo-C-\frac{d}{2}+1}{C-m_--1} \qquad
	\mu_2=\frac{\mpo-C+\frac{d}{2}-1}{C-m_-+1} \OCTC{\qquad}{\cr}
	\mu_3=\frac{1+\mu_1+\mu_2}{C+m_-}
	\end{equation}
Useful relations involving these constants are contained in the appendix.

The homogeneous version of \eqref{12.2} is exactly the same as \eqref{5.2} but 
with $\frac{d}{2}$ and $C$ replaced by
$\frac{d}{2}+1$ and $m_-$, respectively. Therefore our solution is \eqref{10.3} with the same changes:
	\begin{equation}
	\label{14.1a}
	\tilde{\psi_0}^H=(kx^0)^{\frac{d+3}{2}}
	\left[
	i\kdgok K_{m_- + \frac{1}{2}}
	+ K_{m_--\frac{1}{2}}
	\right] c_0^+(\kvec)
	\end{equation}
The complete solution for $\tilde{\psi}_0$ is just the sum of the two parts \eqref{14.1} and \eqref{14.1a}.

To find $\tilde{\psi}_i$  we use the $a=i$ components of \eqref{12.1}:
	\begin{equation}
	\label{15.1}
	\left[x^0\gamma_0{\partial}_0 - i x^0 \kdg - \frac{d}{2} \gamma_0 - m_-\right]\tilpsii \TC{\cr}
	= {\gamma_i}\tilde{\psi}_0+\left[\half\gamma_0{\gamma_i}
		-ix^0 k_i - \mpo{\gamma_i}\right]\gamma \cdot \tilde{\psi}
	\end{equation}
On the RHS of \eqref{15.1}, we have terms from \eqref{14.1}, \eqref{14.1a} and \eqref{10.3}. These terms all consist
 of some power of $kx^0$
and a Bessel function of order $C \pm \half$ or $m_- \pm \half$.
We consider $\tilpsii$ in three parts: $\tilpsii=\tilpsii^H+\tilpsii^C+\tilpsii^{m_-}.$
The homogeneous equation is once again the same as \eqref{5.2} 
with the replacement $C \rightarrow m_-$, so we have
	\begin{equation}
	\label{22.1}
	\tilpsii^H=(kx^0)^\frac{d+1}{2}
	\left[
	i\kdgok K_{m_-+\half}
	+ K_{m_--\half}
	\right]{b^+_i}(\kvec)
	\end{equation}
where $b^+_i(\kvec)$ is free.
Re-using our $A$ parameters in \eqref{9.1}, we make the ansatz
	\begin{equation}
	\tilpsii^C=(kx^0)^\frac{d+1}{2}\LBS\big(A^{(1)}_i + (kx^0) A^{(2)}_i \big)K_{C+\half} \TC{\cr}
		+ ( A^{(3)}_i + (kx^0) A^{(4)}_i )K_{C-\half}\RBS
	\end{equation}
Evaluating \eqref{9.1} with $P=m_-,q=C,n=\frac{d}{2},l=\frac{d+1}{2}$ and matching the result with the corresponding 
Bessel functions on the RHS of \eqref{15.1} we obtain five equations for the four parameters, but they are consistent,
and the result is
	\begin{eqnarray}
	\label{19.1}
	\tilpsii^C &=& (kx^0)^\frac{d+1}{2}
	\TC{\nonumber\\&\times&} \Bigg\{
	K_{C+\half}\left[-i\frac{\mu_2-\mpo+\half}{C+m_-}{\gamma_i} - \OC{(}kx^0\OC{)}\mu_3\kiok\right] \kdgok b_0^+
		 \nonumber\\
	&+& K_{C-\half}\left[\frac{\mu_1+\mpo+\half}{C+m_-}{\gamma_i} + i \OC{(}kx^0\OC{)}\mu_3\kiok\right] b_0^+
	\Bigg\} 
	\end{eqnarray}
Exactly the same procedure, using the ansatz
	\begin{equation}
        \tilpsii^{m_-}=(kx^0)^\frac{d+1}{2}
	\LBS \big(A^{(1)}_i + (kx^0) A^{(2)}_i \big)K_{m_-+\half} \TC{\cr}
		{}+ \big( A^{(3)}_i + (kx^0) A^{(4)}_i \big)K_{m_--\half} \RBS
	\end{equation}
yields the solution
	\begin{eqnarray}
	\label{21.1}
	\tilpsii^{m_-} &=&(kx^0)^\frac{d+1}{2}
	\Bigg\{
	K_{m_-+\half}
		\LBS
		i\kdgok c^+_i (\kvec) \TC{\nonumber\\&&} - (2m_-+1)\kiok\kdgok c_0^+ 
		+{\gamma_i} c_0^+ + i (kx^0) \kiok c_0^+ \RBS \nonumber\\
	&+& K_{m_--\half}\left[c^+_i (\kvec) - (kx^0) \kiok \kdgok c_0^+\right]
	\Bigg\}
	\end{eqnarray}
where $c^+_i (\kvec)$ is free.

At this point we notice that when \eqref{22.1}, \eqref{19.1}, and \eqref{21.1} are combined,
the free quantities $c^+_i$ and $b^+_i$ always appear together as $c^+_i + b^+_i$; we 
thus lose no generality in choosing $c^+_i=0$.

Lastly, for the entire solution to be consistent, we require that $\gamma_0\tilde{\psi}_0+{\gamma_i} \tilpsii$ 
calculated from \eqref{14.1}+\eqref{14.1a} and \eqref{22.1}+\eqref{19.1}+\eqref{21.1} be equal to the same
quantity $\gamma\cdot\tilde{\psi}$ given
by \eqref{10.3}. Equating the two gives a formula
	\begin{equation}
	c_0^+=i\frac{1+\mu_1}{m_1} \kdbok 
	\end{equation}
and also a condition on the otherwise free $b^+_i$ 
	\begin{equation}
	\label{gdotxi}
	\gdb = 0
	\end{equation}

The complete solution for the field $\tilde{\psi}_a$ is thus given by \eqref{14.1} and \eqref{14.1a},
	\begin{eqnarray}
	\label{24.1}
	\tilde{\psi_0}&=&(kx^0)^\frac{d+1}{2}
	\Bigg\{
	K_{C+\half}\left[i \mu_2 \kdgok  + (kx^0)\mu_3\right]b_0^+ \OCTC{+}{\nonumber\\&+&}
	 K_{C-\half}\left[-\mu_1 + i(kx^0)\mu_3\kdgok\right]b_0^+ \OC{\nonumber\\&&} + kx^0
	 \TC{\nonumber\\&\times&} \left[ - \kdgok K_{m_- + \frac{1}{2}} + i K_{m_- - \frac{1}{2}}\right]\frac{1+\mu_1}{m_1}\kdbok
	\Bigg\}
	\end{eqnarray}
and by \eqref{22.1}, \eqref{19.1}, and \eqref{21.1}:
	\begin{eqnarray}
	\label{24.2}
	\lefteqn{ \tilpsii=(kx^0)^\frac{d+1}{2}
	    \Bigg\{
		K_{C+\half}
		\left[
		i\frac{1+\mu_2}{d} 
		{\gamma_i}
		- \OC{(}kx^0\OC{)}\mu_3\kiok
		\right] \kdgok b_0^+ } \nonumber\\
	&& {}+K_{C-\half}
		\left[
		\frac{1+\mu_1}{d} 
		{\gamma_i}
		+i (kx^0)\mu_3\kiok
		\right] b_0^+ \nonumber\\
	&& {}+K_{m_-+\half}\LBS i\kdgok b^+_i + \LBR i\gamma_i - i(2m_-+1) \kdgok \kiok 
		\TC{\nonumber\\&&}-(kx^0)\kiok
		\RBR
		\frac{1+\mu_1}{m_1}\kdbok 
		\RBS \nonumber\\
	&& {}+K_{m_--\half}\left[b^+_i  - i (kx^0) \frac{1+\mu_1}{m_1} \kiok \kdgok \kdbok  \right] 
	\Bigg\}
	\end{eqnarray}
where $b^+_0(\kvec)$ is free, and $b^+_i(\kvec)$ are subject only to \eqref{gdotxi}.
It can be seen that the solutions found in \cite{Corley}, \cite{Volovich}, and \cite{KoshelevRitchkov} 
are special cases of the above solution.

The form of the solution to the conjugate equation \eqref{1.1conj} may be found by conjugating this result. Care must be taken when defining the `bar' operation. We define it in the following way,
	\begin{equation}
	\label{bar}
	\overline{X} \equiv X^\dagger\big\arrowvert_{m \rightarrow -m}
	\end{equation}
where $m \rightarrow -m$ means that we change the sign of both $m_1$ and $m_2$. Under this operation,
 $m_-\rightarrow -m_-$, $m_+\rightarrow -m_+$, $C \rightarrow -C$,
 $\mu_1 \leftrightarrow \mu_2$, and $\mu_3 \rightarrow -\mu_3$.
It is of special importance to realise that $\psi$ and $\overline{\psi}$ are 
independent quantities; taking the conjugate of $\psi$ in this
way only gives the form of the solution for $\overline{\psi}$, and the resulting arbitrary functions 
$\overline{b}_i(\kvec)$ and $\overline{b}_0(\kvec)$ will be unrelated to $b_i$ and $b_0$. In all other cases, the operation
of conjugation will produce not independent quantities, but conjugated ones. 
We will make frequent use of this notation in the remainder of this paper.


\section{Adding a Surface Term to the Action}
\label{SURFsection}
\newcommand{\kint}{\int\frac{\textup{d}^dk}{(2\pi)^d}e^{-i\kvec\cdot(\xvec-\yvec)}}
Along the lines of \cite{Henneaux} we vary the action \eqref{RSaction} and examine surface 
terms which do not vanish when the equations of motion \eqref{1.1}, \eqref{1.1conj} are satisfied. This 
has been done in \cite{Rashkov} and the result is that the term we need to add to 
the action is, in the notation of \cite{Henneaux},
	\begin{equation}
	\label{AddAct}
	C_\infty=\half \int \textup{d}^dx\left(\overline{\psi}_{i(0)}\psi_{i(0)} \alphasign 
		\overline{\psi}_{i(0)}\gamma_i\gamma_j\psi_{j(0)}\right)
	\end{equation}
Thus when we insert the classical solution into the action, only this 
surface term remains and it can be written in the form
	\begin{equation}
	\label{SurfAct}
	I=\half \int\frac{\textup{d}^dk}{(2\pi)^d} \LBR\overline{\psi}_{i(0)}({\bf k})
			\psi_{i(0)}(-{\bf k}) \TC{\cr} \alphasign 
		\overline{\psi}_{i(0)}({\bf k})\gamma_i\gamma_j\psi_{j(0)}(-{\bf k}) \RBR
	\end{equation}
We shall use this in the next section to calculate the correlators. This amounts to doing the 
$k$-integral in \eqref{SurfAct} and taking the $\epsilon \rightarrow 0$ limit. As observed in
\cite{MuckVish1}, this must be done with care by formulating a Dirichlet boundary value problem
not simply at $x^0=0$ but at $\epsilon$ and taking the limit at the end. 

\section{Boundary CFT correlator}
\label{CORRsection} 
\newcommand{\Kcp}{K_{C+\frac{1}{2}}}
\newcommand{\Kcm}{K_{C-\frac{1}{2}}}
\newcommand{\Kmp}{K_{m_-+\frac{1}{2}}}
\newcommand{\Kmm}{K_{m_--\frac{1}{2}}}
\newcommand{\kh}{\kdgok}  
\newcommand{\kdb}{{\bf k} \cdot {\bf b}}

 In the case of a spinor field \cite{MuckVish1}, it was found that as the boundary
at $x^0= \epsilon \rightarrow 0$ is approached, $\psi^+$ and $\psi^-$ are related by a factor of 
some power of $\epsilon$ with the consequence that one may be specified on the boundary, while the 
other must vanish. Since we will find the same behaviour in the present case, we split the field 
into two parts, $\psi^+$ and $\psi^-$. We now set about inverting
\eqref{24.2} to write the parameters $b_i$ and $b_0$ in terms of the boundary fields
 $\psi_i(k\epsilon)$, which we will abbreviate as $\psi_{i\epsilon}$. From \eqref{24.2} we have
	\begin{eqnarray}
	\label{B1.1}
\OCTC{
	\lefteqn{\psi^+_{i\epsilon} = (k \epsilon)^\frac{d+1}{2}\bigg\{
		i \left[ \Kcp \frac{1+\mu_2}{d} \gamma_i \kh  
		+ k \epsilon \Kcm \mu_3 \kiok \right] b_0^+ } 
		\nonumber\\ 
		&& {}+ \Kmm b_i^+ - k \epsilon \Kmp \frac{1+\mu_1}{m_1} \kiok \kdbok \bigg\}
}{
	\psi^+_{i\epsilon} &=& (k \epsilon)^\frac{d+1}{2}\bigg\{
		i \bigg[ \Kcp \frac{1+\mu_2}{d} \gamma_i \kh \nonumber\\ 
		&+& k \epsilon \Kcm \mu_3 \kiok \bigg] b_0^+  
		\nonumber\\ 
		&+& \Kmm b_i^+ - k \epsilon \Kmp \frac{1+\mu_1}{m_1} \kiok \kdbok \bigg\}
}	\end{eqnarray}
	\begin{eqnarray}
	\label{B1.8}
\OCTC{
	\lefteqn{\psi^-_{i\epsilon} = (k \epsilon)^\frac{d+1}{2}\bigg\{
		\left[\Kcm\frac{1+\mu_1}{d}\gamma_i - k\epsilon\Kcp\mu_3\kiok\kdgok \right]b_0^+
			+i\Kmp\kdgok b_i^+ } \nonumber\\ 
		&& {}+ \left[ \Kmp\left(\gamma_i-(2m_-+1)\kdgok\kiok \right)
		-k\epsilon\Kmm\kiok\kdgok\right]i\frac{1+\mu_1}{m_1}\kdbok\bigg\}
}{
	\psi^-_{i\epsilon} &=& (k \epsilon)^\frac{d+1}{2} \nonumber\\
		&\times&\bigg\{
		\left[\Kcm\frac{1+\mu_1}{d}\gamma_i - k\epsilon\Kcp\mu_3\kiok\kdgok \right]b_0^+ \nonumber\\
		&+&i\Kmp\kdgok b_i^+  \nonumber\\ 
		&+& \LBS \Kmp\left(\gamma_i-(2m_-+1)\kdgok\kiok \right) \nonumber\\
		&-&k\epsilon\Kmm\kiok\kdgok\RBS i\frac{1+\mu_1}{m_1}\kdbok\bigg\}
}
	\end{eqnarray}
Multiplying \eqref{B1.1} from the left by $\gamma_i$ and alternately by $\kiok$ gives two equations which
can be solved simultaneously for $b_0$ and $\kdb$ in terms of $\psi_{i\epsilon}$.
We note that in contrast to \cite{KoshelevRitchkov}, where a similar parameter $b_0$
cannot be determined in terms of the boundary data and is arbitrarily set to zero, $b_0$ here
can be written in terms of the boundary field.
Substituting $b_0$ back into \eqref{B1.1} now allows us to solve for $b_i$. 
Inserting these expressions for $b_0$, $\kdb$, 
and $b_i$ into \eqref{B1.8}, $\psi^-_{i\epsilon}$ can be expressed in terms of $\psi^+_{i\epsilon}$. 
Since the $b$ parameters are eliminated in this process, we are 
not free to impose any restrictions on them.
The result is
	\begin{equation}
	\label{B3.4}
	\psi^-_{i\epsilon} = O_{ij}\psi^+_{j\epsilon}
	\end{equation}
where
	\begin{equation}
	\label{B3.2}
	O_{ij}=f_1\gamma_i\kdgok\gamma_j+f_2\kiok\gamma_j+f_3\gamma_i\kjok+f_4\kiok\kdgok\kjok \TC{\cr}
			+f_5\kdgok\delta_{ij}
	\end{equation}
and the each $f$ is a purely imaginary function of $(k\epsilon)$,
		\newcommand{\fI}{\left[\Kcm\Kmm\frac{1+\mu_1}{d}-\Kcp\Kmp\frac{1+\mu_2}{d}
					\right]}
		\newcommand{\fII}{\left[\Kmm-k\epsilon\Kmp\frac{1+\mu_1}{m_1}\right]}
		\newcommand{\fIII}{\Kmp\Kmm\frac{1+\mu_1}{m_1}}
		\newcommand{\fIIIa}{\Kmp\frac{1+\mu_1}{m_1}}
		\newcommand{\fIV}{\left[\Kcp\frac{1+\mu_2}{d}+k\epsilon\Kcm\mu_3\right]}
		\newcommand{\fV}{\LBS 2\Kcp\Kmp\frac{1+\mu_2}{d} \OCTC{+}{\\&+&} k\epsilon\Kcm\Kmp\mu_3
					-k\epsilon\Kcp\Kmm\mu_3\RBS}
		\newcommand{\fVI}{\frac{1+\mu_1}{m_1}\LBS k\epsilon\Kmp^2-k\epsilon\Kmm^2
					\OCTC{-}{\\&-&}(2m_-+1)\Kmp\Kmm\RBS}
		\newcommand{\fVII}{\left[\Kcp(1+\mu_2)+k\epsilon\Kcm\mu_3\right]}
{ \OCTC{\footnotesize}{}  
	\begin{eqnarray*}
	f_1 & = & \frac{1}{D}\fI \TC{\\&\times&} \fII \\
	    & + & \frac{1}{D}\fIII \TC{\\&\times&} \fIV	
	\end{eqnarray*}
	\begin{eqnarray*}
	f_2 & = & \frac{1}{D}\fV\TC{\\&\times&}\fII 	\\
	    & + & \frac{1}{D}\fVI\TC{\\&\times&}\fIV 	
	\end{eqnarray*}
	\begin{eqnarray*}
	f_3 & = & \frac{1}{D}\fI \TC{\\&\times&}k\epsilon\fIIIa \OCTC{\\ & - & }{-} \frac{1}{D}\fIII\TC{\\&\times&}\fVII 	 
	\end{eqnarray*}
	\begin{eqnarray*}
	f_4 & = & \frac{1}{D}\fV \TC{\\&\times&}k\epsilon\fIIIa \\
	    & - & \frac{1}{D}\fVI\TC{\\&\times&}\fVII 	
	\end{eqnarray*}
}
	\begin{equation}
	\label{B3.3}
	f_5  =  i \frac{\Kmp}{\Kmm} 
	\end{equation}
with the denominator given by 
	\begin{equation}
	D=i\LBS \Kmm^2\Kcp(1+\mu_2)+k\epsilon\Kmm^2\Kcm\mu_3 \TC{\cr}
	-k\epsilon\Kmp\Kmm\Kcp\mu_3\RBS
	\end{equation}

In exactly the same way, we may also express $\psi^+_{i\epsilon}$ in terms of $\psi^-_{i\epsilon}$, with the result
	\begin{equation}
	\label{B6.4}
	\psi^+_{i\epsilon} = Q_{ij}\psi^-_{j\epsilon}
	\end{equation}
where
	\begin{equation}
	\label{B15.4}
	Q_{ij}=\overline{f_1}\gamma_i\kdgok\gamma_j+\overline{f_2}\kiok\gamma_j+\overline{f_3}\gamma_i\kjok
		+\overline{f_4}\kiok\kdgok\kjok \TC{\cr} +\overline{f_5}\kdgok\delta_{ij}
	\end{equation}
Conjugating \eqref{B3.4} and \eqref{B6.4} gives us the additional relations
	\begin{equation}
	\label{Obar}
	\overline{\psi}^-_{i\epsilon} = \overline{\psi}^+_{j\epsilon}\overline{O}_{ij}	
	\end{equation}
	\begin{equation}
	\label{Qbar}
	\overline{\psi}^+_{i\epsilon} = \overline{\psi}^-_{j\epsilon}\overline{Q}_{ij}	
	\end{equation}

It is now possible to write \eqref{SurfAct} in a simple 
form. We will, in \eqref{masscase} below, consider a case in which it is necessary to express 
\eqref{SurfAct} in terms of only $\overline{\psi}^+$ and $\psi^-$. This is easily done by
means of \eqref{B6.4} and \eqref{Obar}. We break up the field into $+$ and $-$ pieces
	\begin{eqnarray}
	\label{Surf22a}
	I &=&\frac{\epsilon^{d+1}}{2} \int\frac{\textup{d}^dk}{(2\pi)^d}
		 \Bigg(
	      \overline{\psi}^+_{i\epsilon}({\bf k})\psi^+_{i\epsilon}(-{\bf k}) 
	      +\overline{\psi}^-_{i\epsilon}({\bf k})\psi^-_{i\epsilon}(-{\bf k}) \nonumber\\
	  &\alphasign& \overline{\psi}^+_{i\epsilon}({\bf k})\gamma_i\gamma_j\psi^+_{j\epsilon}(-{\bf k}) 
	      \alphasign \overline{\psi}^-_{i\epsilon}({\bf k})\gamma_i\gamma_j\psi^-_{j\epsilon}(-{\bf k}) 
		\Bigg) \nonumber\\
	  &=&\frac{\epsilon^{d+1}}{2} \int \textup{d}^dx\textup{d}^dy \left(\overline{\psi}^+_{i\epsilon}
			({\bf x})
			\Omega_{ij}({{\bf x}-{\bf y}})
			\psi^-_{j\epsilon}({\bf y}) \right)
	\end{eqnarray}
and then write the correlator as
	\begin{equation}
	\label{Surf22b}
	\Omega_{ij}({{\bf x}-{\bf y}})=\kint \TC{\cr\times} \left[\gamma_i\gamma_lQ_{lj}+\overline{O}_{li}\gamma_l\gamma_j
			\alphasign Q_{ij} \alphasign \overline{O}_{ji} \right]
	\end{equation}
The formula which will be used for this integral\footnote{We will see later on that we have no need 
of local terms; we do not include here terms which contribute only when ${\bf x}$=0.}, properly 
regularised \cite{MuckVish4}, is
	\begin{equation}
	\label{IntFormula}
	\int \frac{\textup{d}^dk}{(2\pi)^d}e^{-i{\bf k}\cdot{\bf x}}k^q=\frac{2^q\Gamma\left(\frac{d+q}{2}\right)}{\pi^{d/2}\Gamma\left(\frac{-q}{2}\right)}\frac{1}{|{\bf x}|^{d+q}}
	\end{equation}

To perform the $k$-integral in \eqref{Surf22b}, we must expand $O$ and $Q$
(and thus each $f$) and determine the power of both $k$ and $\epsilon$ in each term.
Firstly, the terms containing an even, non-negative power of $k$ must be discounted, since
they will vanish by the integral formula \eqref{IntFormula}. Factors of the form $k_i$ 
do not pose a problem here; such a factor can be converted to $i\partial^{(x)}_i$, taking it outside the 
integral. Secondly, we must keep
only leading-order terms in $\epsilon$ of what remains.
To this end we introduce the notation
	\begin{equation}
	\label{B16.2}
	G_{\alpha\beta\gamma} \equiv
		 K_{C+\alpha}(k\epsilon)K_{m_-+\beta}(k\epsilon)K_{m_-+\gamma}(k\epsilon)
	\end{equation}	
and since we will always deal with $\alpha,\beta,\gamma=\pm 1/2$, we abbreviate this further in an obvious way.
The first four $f$ functions may now be written as
\OCTC{
	\begin{eqnarray}
	\label{B16.3}
	f_1 &=& \frac{1}{D}\big[C_1G_{---}+C_2G_{++-}+C_3G_{-+-}k\epsilon+C_4G_{+++}k\epsilon \big]\nonumber\\
	f_2 &=& \frac{1}{D}\big[C_5G_{++-}+C_6G_{-+-}k\epsilon+C_7G_{+--}k\epsilon \nonumber\\
	    &+& C_8G_{+++}k\epsilon+C_{10}G_{++-}(k\epsilon)^2 +C_{11}G_{---}(k\epsilon)^2 \big] \nonumber\\
	f_3 &=& \frac{1}{D}\big[C_{12}G_{-+-}k\epsilon+C_{13}G_{+++}k\epsilon+C_{14}G_{++-} \big]\nonumber\\
	f_4 &=& \frac{1}{D}\big[C_{15}G_{+++}k\epsilon
		+C_{17}G_{++-}(k\epsilon)^2
		+C_{18}G_{+--}k\epsilon \nonumber\\
	    &+& C_{19}G_{++-}+C_{20}G_{---}(k\epsilon)^2+C_{21}G_{-+-}k\epsilon\big]
	\end{eqnarray}		
}{
	\begin{eqnarray*}
	f_1 &=& \frac{1}{D}\big[C_1G_{---}+C_2G_{++-} \\
	    &+&  C_3G_{-+-}k\epsilon+C_4G_{+++}k\epsilon \big]
	\end{eqnarray*}
	\begin{eqnarray*}
	f_2 &=& \frac{1}{D}\big[C_5G_{++-}+C_6G_{-+-}k\epsilon+C_7G_{+--}k\epsilon \\
	    &+& C_8G_{+++}k\epsilon+C_{10}G_{++-}(k\epsilon)^2 +C_{11}G_{---}(k\epsilon)^2 \big]
	\end{eqnarray*}
	\begin{eqnarray*}
	f_3 &=& \frac{1}{D}\big[C_{12}G_{-+-}k\epsilon+C_{13}G_{+++}k\epsilon+C_{14}G_{++-} \big]
	\end{eqnarray*}
	\begin{eqnarray}
	\label{B16.3}
	f_4 &=& \frac{1}{D}\big[C_{15}G_{+++}k\epsilon
		+C_{17}G_{++-}(k\epsilon)^2
		+C_{18}G_{+--}k\epsilon \nonumber\\
	    &+& C_{19}G_{++-}+C_{20}G_{---}(k\epsilon)^2+C_{21}G_{-+-}k\epsilon\big]
	\end{eqnarray}		
}
where $C_{1\dots 21}$ are constants which can be read off from \eqref{B3.3}.
Only the following will be needed
	\begin{equation}
	\label{B17.1}
	C_1 = \frac{1+\mu_1}{d} \qquad\qquad
	C_2 = \frac{1+\mu_2}{d}\left(\frac{1+\mu_1}{m_1}-1\right) \cr
	C_{18} = -C_5 = -C_{14} = \frac{(1+\mu_1)(1+\mu_2)}{m_1} \OCTC{\qquad\qquad}{\cr}
	C_{19} = (2m_-+1)C_{18}
	\end{equation}

Now we use the small-argument expansion of the modified Bessel function
	\begin{equation}
	\label{bessel}
	\TC{2}K_\nu(z)=\OC{\frac{1}{2}}\Gamma(-\nu)\left(\frac{z}{2}\right)^\nu(1+\dots)
		+\OC{\frac{1}{2}}\Gamma(\nu)\left(\frac{z}{2}\right)^{-\nu}(1+\dots)
	\end{equation}
where the dots indicate successive even powers of $z$. Clearly, which term we consider
to be leading-order depends on the order of the Bessel function. 
To settle this point, from now on we turn our attention to the specific case
	\begin{equation}
	\label{masscase}
	m_->\frac{1}{2} \qquad\qquad C<-\frac{1}{2}
	\end{equation}
Other cases may be considered in a similar fashion. A quick inspection of the $f$ functions shows
that in this case, it is $\psi^-$ which may be specified on the boundary, so we will need
to make use of \eqref{B15.4} and \eqref{Obar}, as mentioned above \eqref{Surf22a}.

We apply \eqref{bessel} to obtain the leading-order term in the denominator $\overline{D}$,
	\begin{equation}
	 \label{Dbar}
	 \overline{D}\doteq\frac{i}{8}M\left(\frac{k\epsilon}{2}\right)^{C-2m_--\frac{3}{2}}
	\end{equation}
where $M=-(1+\mu_1)\Gamma(1/2-C)\Gamma(1/2+m_-)^2$. Here we have introduced the dotted 
equal sign $\doteq$ which denotes equality up to leading order in $\epsilon$, 
discounting terms which vanish when integrated due to their power of $k$, as explained 
above \eqref{B16.2}.\footnote{In the case of \eqref{Dbar}, the $\doteq$ functions in only the first way
 since $\overline{D}$ is not integrated by itself.}

Expanding the $\kvec$-dependent part of a general term from $\overline{O}$ or $Q$, we find
	\newcommand{\keot}{\left(\frac{k\epsilon}{2}\right)}
	\begin{eqnarray}
	\label{B20.1}
\OCTC{	\frac{\overline{G_{\alpha\beta\gamma}}(k\epsilon)^P}{\overline{D}k^l}&\doteq& 
		\frac{-i2^P}{Mk^l}\left[\Gamma(\alpha-C)\keot^{\half-\alpha+P}
			+\Gamma(C-\alpha)\keot^{\half-2C+\alpha+P}\right] \nonumber\\
		&\times&\left[\Gamma(\beta-m_-)\keot^{\half+2m_- - \beta}
			+\Gamma(m_--\beta)\keot^{\half+\beta}\right] \nonumber\\
		&\times&\left[\Gamma(\gamma-m_-)\keot^{\half+2m_- - \gamma}
			+\Gamma(m_--\gamma)\keot^{\half+\gamma}\right]
}{	\lefteqn{\frac{\overline{G_{\alpha\beta\gamma}}(k\epsilon)^P}{\overline{D}k^l}\doteq
					\frac{-i2^P}{Mk^l}\keot^{P+\half}} \\
		&\times&\left[\Gamma(\alpha-C)\keot^{-\alpha}
			+\Gamma(C-\alpha)\keot^{-2C+\alpha}\right] \nonumber\\
		&\times&\left[\Gamma(\beta-m_-)\keot^{2m_- - \beta}
			+\Gamma(m_--\beta)\keot^{\beta}\right] \nonumber\\
		&\times&\left[\Gamma(\gamma-m_-)\keot^{2m_- - \gamma}
			+\Gamma(m_--\gamma)\keot^{\gamma}\right] \nonumber
}	\end{eqnarray}
In the expansion of this product, we will refer to individual terms by the signs of $\alpha$, $\beta$, 
and $\gamma$ in the exponent. By inspection it is seen that the $-\alpha+\beta+\gamma$ term 
is leading-order in $\epsilon$; therefore 
we should ask whether this term will vanish when we do
the $\kvec$-integral.

\OC{{	\begin{table}[!hbp]
	\caption{Powers of $k$ and $\epsilon$ in each instance of the leading-order term in \eqref{B20.1}}
	\label{B20.2}
	\center			\begin{tabular}{ccc|cc}
				$l$ & $\alpha\beta\gamma$ & $P$ & Power of $k$ & Power of $\epsilon$ \\
				\hline
				$1$ & $---$ & $0$ & $0$ & $1$ \\
				$1$ & $++-$ & $0$ & $0$ & $1$ \\
				$1$ & $-+-$ & $1$ & $2$ & $3$ \\
				$1$ & $+++$ & $1$ & $2$ & $3$ \\
				$1$ & $+--$ & $1$ & $0$ & $1$ \\
				$1$ & $++-$ & $2$ & $2$ & $3$ \\
				$1$ & $---$ & $2$ & $2$ & $3$ \\
				$3$ & $+++$ & $1$ & $0$ & $3$ \\
				$3$ & $++-$ & $2$ & $0$ & $3$ \\
				$3$ & $+--$ & $1$ & $-2$ & $1$ \\
				$3$ & $++-$ & $0$ & $-2$ & $1$ \\
				$3$ & $---$ & $2$ & $0$ & $3$ \\
				$3$ & $-+-$ & $1$ & $0$ & $3$ \\
				\end{tabular}
	\end{table}}}
\TC{{ \center \hrule \vspace{1pt} \hrule \vspace{6pt} 
	 TABLE I. Powers of $k$ and $\epsilon$ in each instance of the leading-order term in \eqref{B20.1} \\
				\begin{tabular}{ccc|cc}
				$l$ & $\alpha\beta\gamma$ & $P$ & Power of $k$ & Power of $\epsilon$ \\
				\hline
				$1$ & $---$ & $0$ & $0$ & $1$ \\
				$1$ & $++-$ & $0$ & $0$ & $1$ \\
				$1$ & $-+-$ & $1$ & $2$ & $3$ \\
				$1$ & $+++$ & $1$ & $2$ & $3$ \\
				$1$ & $+--$ & $1$ & $0$ & $1$ \\
				$1$ & $++-$ & $2$ & $2$ & $3$ \\
				$1$ & $---$ & $2$ & $2$ & $3$ \\
				$3$ & $+++$ & $1$ & $0$ & $3$ \\
				$3$ & $++-$ & $2$ & $0$ & $3$ \\
				$3$ & $+--$ & $1$ & $-2$ & $1$ \\
				$3$ & $++-$ & $0$ & $-2$ & $1$ \\
				$3$ & $---$ & $2$ & $0$ & $3$ \\
				$3$ & $-+-$ & $1$ & $0$ & $3$ \\
				\end{tabular} \vspace{1pt}\hrule\vspace{1pt}\hrule\vspace{6pt} }}
\OCTC{Table~\ref{B20.2}}{Table~I} shows on the LHS all instances, in \eqref{B16.3}, of the general term \eqref{B20.1}.
It should be noted that there are also factors of $k^{-1}$ which come from \eqref{B3.2} and \eqref{B15.4} so
that, for example, we should consider $\frac{\overline{f_4}}{k^3}$ rather than just $\overline{f_4}$.
On the RHS are the resulting powers of $\epsilon$ and $k$ 
in the leading-order term in \eqref{B20.1}. 
We see that all elements in the table will 
vanish when we integrate over $\kvec$ except the two entries indicating $k^{-2}$. However,
when this part of $\overline{f_4}$ is evaluated by substituting $\overline{C_{18}}$ and $\overline{C_{19}}$
from \eqref{B17.1}, we see that these two terms neatly cancel each other. 
Thus, the leading-order term in \eqref{B20.1}
will always vanish when we integrate. Since it cannot contribute, we must analyse the seven
remaining higher-order terms to see which do.
By similar arguments we see that the next-order terms
in \eqref{B20.1} (for generic $\alpha\beta\gamma$) are the $-\alpha-\beta+\gamma$, $-\alpha+\beta-\gamma$,
and $+\alpha+\beta+\gamma$ terms. 
We will make no assumption as to whether $C$ is larger or smaller in magnitude than $m_-$, 
so we must keep all three of these.\footnote{Actually, the $-\alpha+\beta-\gamma$ term turns out 
not to contribute anyway.} It is also assumed that the masses are not special in that when we 
integrate a term of the form $k^{(masses)}$, it does not vanish. Considering again each instance of
the general term \eqref{B20.1}, we find that the leading-order terms go as $\epsilon^{2m_-}$ 
and $\epsilon^{-2C}$, and that they correspond only to the instances $\alpha\beta\gamma=---$ and
$\alpha\beta\gamma=++-$, both with $P=0$.\footnote{The value of $l$ in \eqref{B20.1} does not matter here since it affects only the power of $k$ which appears, and not $\epsilon$.} Hence, the only terms in \eqref{B16.3}
which will survive the $\epsilon \rightarrow 0$ 
limit and the $\kvec$-integration are the $\overline{C_1}$, $\overline{C_2}$, $\overline{C_5}$,
 $\overline{C_{14}}$, and $\overline{C_{19}}$ terms,
	\begin{equation}
	\label{B22.2}
	\frac{\overline{f_1}}{k} \doteq \overline{C_1}\frac{\overline{G_{---}}}{\overline{D}k}
					+\overline{C_2}\frac{\overline{G_{++-}}}{\overline{D}k} \qquad
	\frac{\overline{f_2}}{k} \doteq \overline{C_5}\frac{\overline{G_{++-}}}{\overline{D}k} \OCTC{\qquad}{\cr}
	\frac{\overline{f_3}}{k} \doteq \overline{C_{14}}\frac{\overline{G_{++-}}}{\overline{D}k} \qquad
	\frac{\overline{f_4}}{k^3} \doteq \overline{C_{19}}\frac{\overline{G_{++-}}}{\overline{D}k^3}
	\end{equation} 
Writing out explicitly from \eqref{B20.1} the terms which will contribute, according to the above 
analysis, we have
	\begin{eqnarray} 
	\label{B22.3}
	\frac{\overline{G_{---}}}{\overline{D}k^l} &\doteq& \frac{-i}{M}
		\Gamma\left(\half+C\right)\Gamma\left(m_--\half\right)^2
		\left(\frac{\epsilon}{2}\right)^{-2C}k^{-2C-l} \nonumber\\
	\frac{\overline{G_{++-}}}{\overline{D}k} &\doteq& \frac{-i}{M}
		\Gamma\left(\half-C\right)\Gamma\left(\half-m_-\right)\Gamma\left(\half+m_-\right)
		\TC{\nonumber\\&\times&}
		\left(\frac{\epsilon}{2}\right)^{2m_-}k^{2m_--1} 
	\end{eqnarray} 	
From \eqref{B3.3}, $\overline{f_5}$ is of order $\epsilon^{2m_-}$, and
	\begin{equation}
	\label{B20a.1}
	\frac{\overline{f_5}}{k} \doteq -i\frac{\Gamma(\half-m_-)}{\Gamma(\half+m_-)}
		\left(\frac{\epsilon}{2}\right)^{2m_-}k^{2m_--1}
	\end{equation} 

\newcommand{\gdp}{{\boldsymbol \gamma}\cdot{\boldsymbol \partial}}
\newcommand{\gdx}{{\boldsymbol \gamma}\cdot|\xvec - \yvec|}
\newcommand{\gdxab}{{\boldsymbol \gamma}_{\alpha\beta}\cdot|\xvec - \yvec|}
\newcommand{\minusMfour}{\frac{1}{\pi^{d/2}}\left(\frac{1+\mu_2}{m_1}+1\right)
	            \frac{\Gamma\left(\frac{d+2m_-+1}{2}\right)}{\Gamma\left(\half+m_-\right)}}
\newcommand{\minusTwoMfour}{\frac{2}{\pi^{d/2}}\left(\frac{1+\mu_2}{m_1}+1\right)
	            \frac{\Gamma\left(\frac{d+2m_-+1}{2}\right)}{\Gamma\left(\half+m_-\right)}}
\newcommand{\Mfive}{\frac{1}{(1+\mu_1)\left(m_--\half\right)^2\pi^{d/2}}
	            \frac{\Gamma\left(\frac{d-2C+1}{2}\right)}{\Gamma\left(\half-C\right)}}
Now the formula \eqref{IntFormula} can be used to find $\int \textup{d} k$ of \eqref{B22.3}. Substituting the results into \eqref{B3.2}
\OCTC{(conjugated) yields \begin{eqnarray}
	\label{B23a.1}
	\lefteqn{\kint\overline{O}_{ij}} \nonumber\\
	&\doteq& \frac{1}{M} \Bigg\{\frac{\overline{C_1}}{2\pi^{d/2}}
		\Gamma\left(m_--\half\right)^2\Gamma\left(\frac{d-2C-1}{2}\right)
		\TC{\nonumber\\&\times&}\gamma_j\gdp\gamma_i
		\frac{\epsilon^{-2C}}{|\xvec-\yvec|^{d-2C-1}} \nonumber\\
	&-&\frac{\overline{C_{19}}}{8\pi^{d/2}\left(\half-m_-\right)}\Gamma\left(\half-C\right)
		\Gamma\left(\half+m_-\right)
		\TC{\nonumber\\&\times&}\Gamma\left(\frac{d+2m_--3}{2}\right)\gdp\partial_i\partial_j
		\frac{\epsilon^{2m_-}}{|\xvec-\yvec|^{d+2m_--3}} \nonumber\\
	&+&\frac{\Gamma\left(\half-C\right)}{2\pi^{d/2}}{\Gamma\left(\frac{d+2m_--1}{2}\right)}
		\TC{\nonumber\\&\times&}\left[\overline{C_2}\gamma_j\gdp\gamma_i+\overline{C_5}\partial_i\gamma_j
		+\overline{C_{14}}\gamma_i\partial_j\right]
		\frac{\epsilon^{2m_-}}{|\xvec-\yvec|^{d+2m_--1}} \nonumber\\
	&+&\frac{\Gamma\left(\frac{d+2m_--1}{2}\right)}{2\pi^{d/2}\Gamma\left(\half+m_-\right)}\gdp
		\frac{\epsilon^{2m_-}}{|\xvec-\yvec|^{d+2m_--1}} \Bigg\}
	\end{eqnarray}
Doing the derivatives and simplifying, we obtain}
{(conjugated), doing the resulting derivatives, and simplifying, we obtain}
	\begin{eqnarray}
	\label{B26.1and27.1}
\OCTC{	\lefteqn{\kint\overline{O}_{ij} \doteq \minusMfour } \nonumber\\
	&\times&\left\{\gdx\left(\delta_{ij}-2\frac{(x-y)_i(x-y)_j}{|\xvec-\yvec|^2}\right)
	+\frac{\gamma_j\gdx\gamma_i}{d}\right\}\frac{\epsilon^{2m_-}}{|\xvec-\yvec|^{d+2m_-+1}} \nonumber\\
	&+& \Mfive \gamma_j\gdx\gamma_i\frac{\epsilon^{-2C}}{|\xvec-\yvec|^{d-2C+1}}
}{
	\lefteqn{\kint\overline{O}_{ij} } \nonumber\\
	& \doteq & \minusMfour \nonumber\\
	&\times& \bigg\{\gdx\left(\delta_{ij}-2\frac{(x-y)_i(x-y)_j}{|\xvec-\yvec|^2}\right) \nonumber\\
	&+& \frac{\gamma_j\gdx\gamma_i}{d}\bigg\}\frac{\epsilon^{2m_-}}{|\xvec-\yvec|^{d+2m_-+1}} \nonumber\\
	&+& \Mfive \nonumber\\
	&\times&\gamma_j\gdx\gamma_i\frac{\epsilon^{-2C}}{|\xvec-\yvec|^{d-2C+1}}
}	\end{eqnarray}
Comparing \eqref{B15.4} with \eqref{B3.2}, we see that it is unnecessary to
calculate $\int \textup{d} k Q_{ij}$
separately; it can be obtained trivially from \eqref{B26.1and27.1} by switching the $i$ and $j$ 
indices in the $\gamma_j\gdx\gamma_i$ terms.

The correlator $\Omega_{ij}(\xvec-\yvec)$ from \eqref{Surf22b} may now be written
	{\setlength\arraycolsep{2pt}
	\begin{eqnarray}
	\label{B26.4and27.2}
	\Omega_{ij}(\xvec-\yvec) &=& M_I
	 \LBS\frac{\gamma_i\gdx\gamma_j}{d} \OCTC{\alphasign}{\nonumber\\&\alphasign&}\gdx
		\left(\delta_{ij}-2\frac{(x-y)_i(x-y)_j}{|\xvec-\yvec|^2}\right)
		\RBS\TC{\nonumber\\&\times&}\frac{\epsilon^{2m_-}}{|\xvec-\yvec|^{d+2m_-+1}} \nonumber\\
	& & {}+M_{I\!I} \gamma_i\gdx\gamma_j\frac{\epsilon^{-2C}}{|\xvec-\yvec|^{d-2C+1}}
	\end{eqnarray}}
Where $M_I$ and $M_{I\!I}$ are constants. The two correlators contained in this expression can
be separated by decomposing the
boundary field $\psi_{i\epsilon}$ into two parts, projecting out the component orthogonal to $\gamma_i$.
To use the AdS/CFT correspondence \eqref{Malda}, we define the boundary 
fields
	\[
	\chi_{(0)} \equiv \gamma_j\psi_j 
	\qquad \textup{and} \qquad
	\psi_{i(0)} \equiv \psi_i-\frac{\gamma_i}{d}\chi_{(0)}
	\]
so that $\gamma_i\psi_{i(0)}=0$. The presence of the spinor field $\chi_{(0)}$ accounts for the 
correct number of degrees of freedom coming from the original $d+1$-dimensional Rarita-Schwinger field.

Rewriting in terms of these new fields, while absorbing appropriate powers of $\epsilon$, gives 
us two correlators, one for the conformal operator coupling to each boundary field
	\begin{eqnarray}
	\label{B30}
	\langle \Pcurlymath{O}_{i\alpha}\overline{\Pcurlymath{O}}_{j\beta} \rangle &=& M_I \gdxab
		\left[\delta_{ij}-2\frac{(x-y)_i(x-y)_j}{|\xvec-\yvec|^2}\right]
		\TC{\nonumber\\&\times&}
		\frac{1}{|\xvec-\yvec|^{d+2m_-+1}} \\
	\label{B31}
	\langle \Pcurlymath{O}'_{\alpha}\overline{\Pcurlymath{O}'}_{\beta} \rangle 
		&=& M_{I\!I} \gdxab\frac{1}{|\xvec-\yvec|^{d-2C+1}}
	\end{eqnarray}
and the scaling dimensions of the operators are $\Delta_{\Pcurlymath{O}_i}=\frac{d}{2}+m_-$ and
$\Delta_{\Pcurlymath{O}'}=\frac{d}{2}-C$.

Each of these correlators is seen to be of the form required by conformal invariance \cite{Zaikov}.
The first correlator \eqref{B30} has been obtained previously
but the condition $\gamma^\mu\psi_\mu=0$ was imposed in \cite{Corley,Volovich}, and
a parameter paralleling $b_0$ set to zero in \cite{KoshelevRitchkov}, with the result that the 
second correlator was not found.
Our construction in section \ref{EOMsection} of the solution to the equations of motion
involves no such restrictions, thus allowing both 
correlators \eqref{B30} and \eqref{B31} to be obtained.
It is interesting to note that since $\Delta_{\Pcurlymath{O}'}$
depends on $C$, which is proportional to $1/m_1$, the limit $m_1 \rightarrow 0$
is not well-defined. Hence, the massless case cannot be recovered in this limit.

\section*{Acknowledgements}
The authors would like to thank W.~M\"uck for helpful discussions. P.~M.~ is grateful
to Simon Fraser University for a Graduate Fellowship. 
This work was supported in part by a grant from NSERC.


\section*{Appendix}
The following are useful identities involving the constants defined in \eqref{14.1mu} and \eqref{10.4}.
	\[
	\mu_1 = \frac{1-d-2m_2}{d-1-2m_-} \qquad
	\mu_2 = \frac{1-d+2m_2}{d-1+2m_-} \OCTC{\qquad}{\cr}
	\mu_3 = \frac{4 m_1 (1-d)}{d(d-1-2m_-)(d-1+2m_-)} \cr
	\frac{\mu_3}{d-1} = \frac{1+\mu_1}{m_1}\frac{1+\mu_2}{d} \OCTC{\qquad\qquad}{\cr}
	\frac{1+\mu_1}{1+\mu_2} = 2m_-\frac{1+\mu_1}{m_1}-1 \cr
	\frac{\mu_1+\mpo+\half}{C+m_-} = \frac{-2m_1}{d(d-1-2m_-)}=\frac{1+\mu_1}{d} \cr
	\frac{\mu_2-\mpo+\half}{C+m_-} = \frac{-2m_1}{d(d-1+2m_-)}=-\frac{1+\mu_2}{d} \cr
	\]




\begin{thebibliography}{10}
\bibitem{Maldacena} J.~Maldacena, \emph{The Large {N} Limit of Superconformal Field Theories and Supergravity}, Adv. Theor. Math. Phys. 2 (1998) 231, hep-th/9711200.
\bibitem{Susskind} L.~Susskind, \emph{The World as a Hologram}, J. Math. Phys 36 (1995) 6377.
\bibitem{GubsKlebPolya} S.~S.~Gubser, I.~R.~Klebanov, and A.~M.~Polyakov, \emph{Gauge Theory Correlators from Non-critical String Theory}, Phys. Lett. \textbf{B428} (1998) 105, hep-th/9802109.
\bibitem{ArefevaVolovich} I.~Y. Aref'eva and I.~V. Volovich, \emph{On Large {N} Conformal Theories, Field Theories in Anti-de~Sitter Space and Singletons}, hep-th/9803028.
\bibitem{Witten} E.~Witten, \emph{Anti-de~Sitter Space and Holography}, Adv. Theor. Math. Phys. 2 (1998) 253, hep-th/9802150.
\bibitem{KlebWitt} I.~R.~Klebanov and E.~Witten, \emph{AdS/CFT Correspondence and Symmetry Breaking}, hep-th/9905104.
\bibitem{MuckVish0} W.~M\"uck and K.~S.~Viswanathan, \emph{Conformal Field Theory Correlators from Classical Scalar Field Theory on Anti-de Sitter Space}, Phys. Rev. \textbf{D58} (1998) 041901, hep-th/9804035.
\bibitem{HenningsonSfestos} M.~Henningson and K.~Sfestos, \emph{Spinors and the AdS/CFT correspondence}, Phys. Lett. \textbf{B431} (1998) 63, hep-th/9803251.
\bibitem{MuckVish1} W.~M\"uck and K.~S.~Viswanathan, \emph{Conformal Field Theory Correlators from Classical Field Theory on Anti-de Sitter Space II. Vector and Spinor Fields}, Phys. Rev. \textbf{D58} (1998) 106006, hep-th/9805145.
\bibitem{DHokerFreeSki} E.~D'Hoker, D.~Z.~Freedman, and W.~Skiba, \emph{Field Theory Tests for Correlators in the AdS/CFT Correspondence}, Phys. Rev. \textbf{D59} (1999) 045008, hep-th/9807098.
\bibitem{Tseytlin} H.~Liu, A.~A.~Tseytlin, \emph{On Four-point Functions in the CFT/AdS Correspondence}, Phys. Rev. \textbf{D59} (1999) 086002, hep-th/9807097.
\bibitem{Corley} S.~Corley, \emph{The Massless Gravitino and the Ads/CFT Correspondence}, Phys. Rev. \textbf{D59} (1999) 086003, hep-th/9808184.
\bibitem{Volovich} A.~Volovich, \emph{Rarita-Schwinger Field in the AdS/CFT Correspondence}, JHEP 9809 (1998) 022, hep-th/9809009.
\bibitem{FreedmanMathur} D.~Z. Freedman, S.~D. Mathur, A. Matusis and L. Rastelli, \emph{Correlation Functions in the $CFT_d/AdS_{d+1}$ Correspondence}, Nucl. Phys. \textbf{B546} (1999) 96-118, hep-th/9804058.
\bibitem{MuckVish2} W.~M\"uck and K.~S.~Viswanathan, \emph{The Graviton in the AdS/CFT Correspondence: Solution via the Dirichlet Boundary Value Problem}, hep-th/9810151.
\bibitem{KoshelevRitchkov} A.~S.~Koshelev and O.~A.~Rytchkov, \emph{Note on the Massive Rarita-Schwinger Field in the AdS/CFT Correspondence}, Phys. Lett. \textbf{B450} (1999) 368-376, hep-th/9812238.
\bibitem{Rashkov} R.~C.~Rashkov, \emph{Note on the Boundary Terms in AdS/CFT Correspondence for Rarita-Schwinger Field}, hep-th/9904098.
\bibitem{Petersen} J.~L.~Petersen, \emph{Introduction to the Maldacena Conjecture on AdS/CFT}, lectures given in Leuven, January 1999, hep-th/9902131.
\bibitem{ArutyunovFrolov} G.~E.~Arutyunov and S.~A.~Frolov, \emph{On the Origin of Supergravity Boundary Terms in the AdS/CFT Corrsepondence}, Nucl. Phys. \textbf{B544} (1999) 576-589, hep-th/9806216.
\bibitem{Henneaux} M.~Henneaux, \emph{Boundary Terms in the AdS/CFT Correspondence for Spinor Fields}, hep-th/9902137.
\bibitem{FerGrilPari} S.~Ferrara, A.~F.~Grillo, G.~Parisi, and R.~Gatto, \emph{The Shadow Operator Formalism for Conformal Algebra. Vacuum Expectation Values and Operator Products}, Lett. Nuovo Cim. 4 (1972) 115.
\bibitem{Zaikov} G.~M.~Sotkov and R.~P.~Zaikov, \emph{Conformal Invariant Two and Three-Point Functions for Fields with Arbitrary Spin}, Rept. on Math. Phys. \textbf{12} (1977) 375.
\bibitem{vanN} H.~J.~Kim, L.~J.~Romans and P.~van~Nieuwenhuizen, \emph{Mass Spectrum of Chiral Ten-Dimensional $N=2$ Supergravity on $S^5$}, Phys. Rev. \textbf{D32} (1985) 389.
\bibitem{Weinberg} S.~Weinberg, \emph{Gravitation and Cosmology}, John Wiley \& Sons, 1972.
\bibitem{MuckVish4} W.~M\"uck and K.~S.~Viswanathan, \emph{A Regularisation Scheme for the AdS/CFT Correspondence}, hep-th/9904039.
\end{thebibliography}
\end{document}